\documentclass[useAMS,usenatbib]{mn2e}
\usepackage{graphicx}
\usepackage{rotating}

\def\la{\raise.5ex\hbox{$<$}\kern-.8em\lower 1mm\hbox{$\sim$}}
\def\ma{\raise.5ex\hbox{$>$}\kern-.8em\lower 1mm\hbox{$\sim$}}

\def\Lsol{L$_{\odot}$ }
\def\kms{$\rm km\, s^{-1}$}
\def\cm3{$\rm cm^{-3}$}
\def\Ts{$\rm T_{*}$~}
\def\Vs{$\rm V_{s}$}
\def\n0{$\rm n_{0}$}
\def\B0{$\rm B_{0}$}

\def\Ne{$\rm N_{e}$~}
\def\Te{$\rm T_{e}$~}

\def\erg{$\rm erg\, cm^{-2}\, s^{-1}$}
\def\mum{$\mu$m~}

\def\L12{L$_{12\mu m}$~}
\def\F12{F$_{12\mu m}$~}
\def\agr{a$_{gr}$}
\def\Hb{H${\beta}$~}
\def\Ha{H${\alpha}$}
\def\Hg{H${\gamma}$~}

\def\RO3{R$_{[OIII]}$}

\title[GRB031203  host]{Shock fronts in the long  GRB031203  host galaxy
} 
\author[M. Contini]{ M. Contini
\\
School of Physics and Astronomy, Tel Aviv University, Tel Aviv
69978, Israel \\
}

\begin{document}


\pagerange{\pageref{firstpage}--\pageref{lastpage}} \pubyear{2009}

\maketitle

\label{firstpage}

\begin{abstract}
The detailed modelling of the spectra observed from the long GRB031203 host galaxy at
different epochs during the 2003-2009 years is presented.  The line profiles show  FWHM  of
 $\sim$ 100 \kms. A  broad line  profile  with  FWHM $\leq$ 400 \kms appears in the line sockets 
 from the 2009 observations. 
We suggest that the narrow  lines  show the   velocity of  star-burst (SB)  debris, while 
the  broad  ones are due to the wind from  SB stars.
The spectra are emitted from the gas downstream of different shock fronts  which are at work on the
edges of the emitting clouds. A head-on-back shock  appears when  the wind from the  SB stars
reaches  the internal edge of the  SB debris moving outwards. A  head-on shock is 
created by  collision of  the  debris   with the ISM clouds. 
Line ratios in both cases are calculated by the coupled effect of  shock and 
photoionization  from the SB.   The models selected by fitting the calculated to the observed 
line ratios  show that  the ionization parameters,  the shock velocities and the gas pre-shock 
densities slowly decrease with time. 
Oxygen metallicities  (12+log(O/H)=8.3-8.48) are lower than solar (8.82) by a factor $<$3 and 
nitrogen metallicities  are lower than solar (12+log(N/H)=8.0) by  factors of 3-5.

\end{abstract}

\begin{keywords}
radiation mechanisms: general --- shock waves --- 
 galaxies: GRB host  --- galaxies: high redshift

\end{keywords}

\section{Introduction}

Long duration gamma-ray bursts (LGRB)  and  short duration GRB (SGRB) have  different origins.
LGRB derive from implosion of a massive star while  SGRB are the product of (neutron star)
NS-NS or NS-BH (black hole) collisions.
LGRBs appear  among young population stars (Fruchter et al 2006, Savaglio et al 2009, etc), while for SGRBs both young 
and old population stars were found (e.g.  Berger 2009,   Savaglio et al. 2009,  etc.).
GRB are distinguished  by  time periods, which are
   evident  from  the distribution of the data.  The  burst period  is T$_{90}$$\leq$ 2s for
SGRB, where T$_{90}$ is the time interval   between 5 percent
to 95 percent gamma-ray photons collected by a given instrument (Kouveliotou et al 1993).
LGRB have longer duration times, up to  several minutes (Kouveliotou et al).
 Fig. 1 shows the dicotomy of the GRB phenomenon.
GRB031203  triggered the  imager on-Board the Integral Satellite (IBIS) instrument  of  the international gamma-ray astrophysics laboratory (INTEGRAL) on 2003 December 3
(Gotz et al 2003)
with a duration of $\sim$ 30s and a peak flux of 1.3 10$^{-7}$ \erg (20-200 keV, Meneghetti \& Gots 2003).
A compact dwarf galaxy coinciding with the X-ray source (Hsia et al 2003)
was later identified as the GRB031203 host at z=0.1055 (Prochaska et al 2004).
A SN event led to the discovery of supernova SN2003lw (Tagliaferri et al 2004).
The X-ray and radio afterglows were soon discovered  at  European Southern Observatory (ESO), starting
near infrared (NIR) observations at the New Technology Telescope (NTT)
7 hours after the GRB and then for the X-ray source in GRB031203.
Guseva et al (2011) reported   the general  claim that "there are no evidences that host galaxies
are not peculiar and are similar to star-forming galaxies in the local and
distant universe". They concluded that for galaxies at z$\geq$0.1, the properties
of the LGRB environment can be retrieved from the ISM of the entire galaxy as  occurs for GRB031203.
The afterglow of GRB031203 was very weak, the faintest ever detected in the optical/NIR.
The detection of the SN optical light implies that an extreme dust obscuration was not the reason for such faintness.

\begin{figure}
\centering
\includegraphics[width=9.2cm]{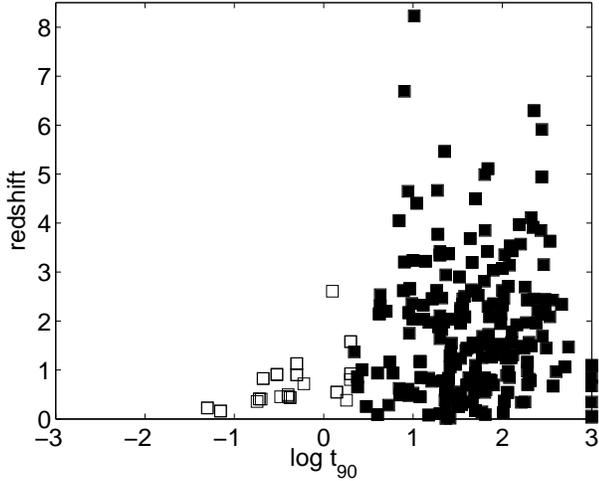}
\caption{Distribution of GRB redshifts versus t$_{90}$. Open squares: SGRB;
 filled squares: LGRB.  Data from Wei et al (2014).
}
\end{figure}

In the latest years we   have been investigating LGRB and  SGRB  host galaxy spectra 
by the detailed calculation of the physical conditions and the element abundances.
The results suggest that LGRB  and  SGRB  host galaxies may be distinguished
by the N/O distribution trend as function of the redshift at low z 
(z$<$4 Contini 2017). Moreover,
it seems that in some of the SGRB hosts (e.g. SGRB100628A) an active galactic nucleus (AGN) can be present and even coupled 
to the star-burst (Contini 2019).
In  this paper we  calculate by  detailed modelling  the  line and continuum spectra of LGRB031203 host galaxy. 
Observations at different epochs were presented by Prochaska et al (2004), Margutti et al (2007), Guseva et al (2011), 
 Watson et al (2011), etc.
Observations of the spectra by Margutti et al (2007) at different   times in the years 2003-2004 
provide the opportunity  to follow the variations of the emitted line ratios observed from GRB031203 host 
on a short time scale. 
 We  focus on  line and continuum  modelling  throughout the host galaxy,
in particular on the physical conditions and metallicities,  on
the contribution to the continuum SED  of  gas bremsstrahlung and reradiation by dust,  on the 
contribution of the underlying old star population in the IR and of
radio synchrotron emission. 
To calculate the physical conditions and the element abundances in LGRB031203 host  we   use
composite models which account for the photoionizing flux from the star-burst  (SB) coupled with shocks.

Afterglow issues,  host physical   features and  metallicities which  characterize GRB031203
 are  summarized in Sect. 2.
The calculations are  presented in Sect. 3 and the
 results of modelling the line ratios observed by Prochaska et al,  Margutti et al (2007), 
Guseva et al (2011) and  Watson et al (2011) 
appear  in Sect. 4. In Sect. 5   the continuum SED observations are compared with 
the results of the  models that were  adopted to fit the line ratios.  Concluding remarks follow in Sect. 6.

\begin{figure*}
\centering
\includegraphics[width=8.8cm]{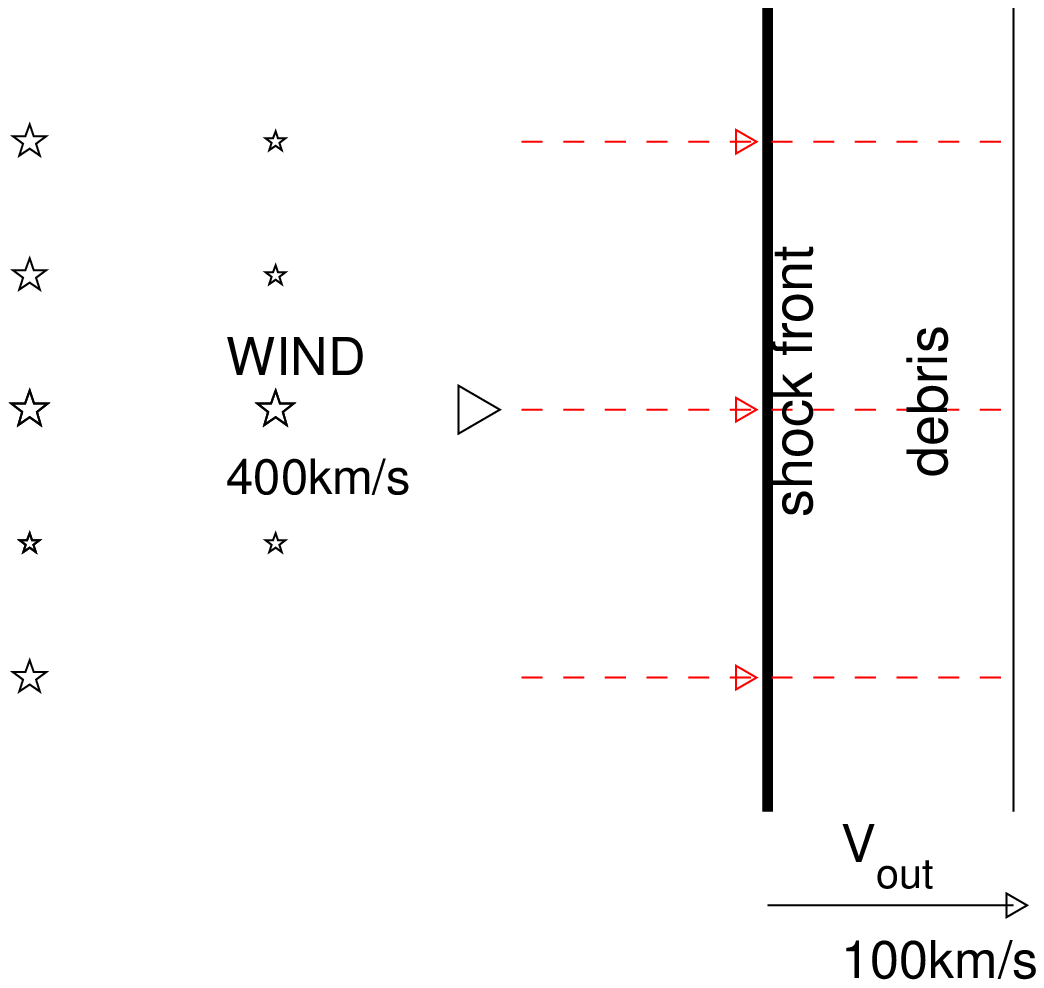}
\includegraphics[width=8.8cm]{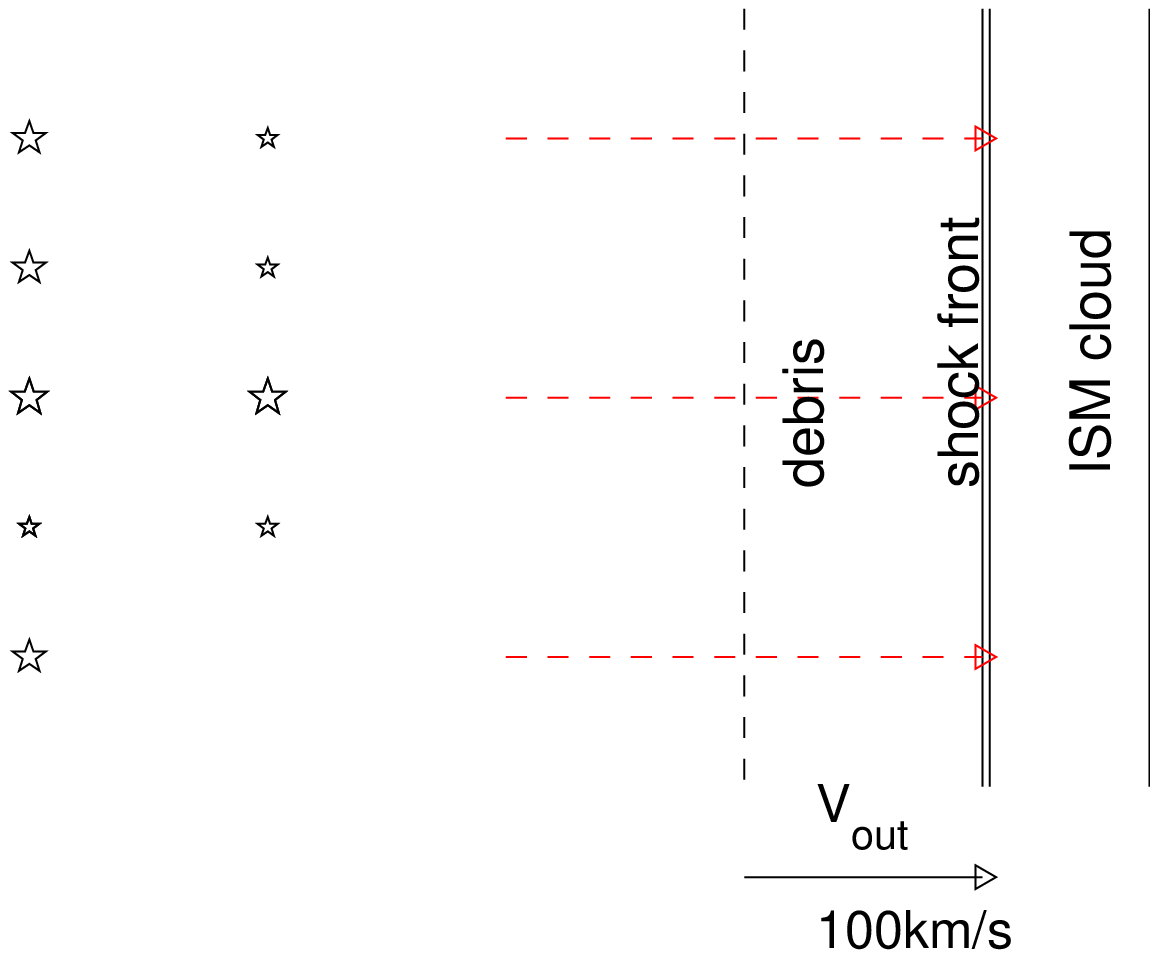}
\caption{Sketch of the head-on (right) and head-on-back (left) shock fronts.
Red dotted lines represent the radiation flux from the SB.
}
\end{figure*}

\section{General GRB  features} 

In the following we summarize some  of the   GRB  issues  which are  significant to  GRB031203,
 in particular,  a weak afterglow, a complex hydrodynamic host field and low metallicities.

\subsection{Afterglow}

An afterglow arises when the  GRB ejecta are slowed down by  collision with circum-burst matter.
It  shows lower energy than the GRB, long lasting  X-ray,  optical emission and a  long-lived
radio emission  due to a  blast wave subrelativistic envelope.
The overall observed isotropic luminosity  is $\sim$ 10$^{51}-10^{52}$ erg s$^{-1}$  (Piran 2004).
 Berger (2009) by the observational evidence suggests a link of LGRB  with  the end of life of massive stars
(Woosley 1993, Woosley \& Bloom 2006) from their association with core collapse supernovae (Hjorth et al 2003).
As a consequence the fireball shell expands (Kobayashi  \& Zhang 2003)
into the pre-burst stellar wind of the progenitor star.  The wind has  a density distribution $\rho$ $\propto$ R$^{-2}$.
The interaction between the shell and the wind  leads to two shocks:  one
 propagating in reverse  into the shell and  the other propagating forward into the wind.
The  velocity of the shock  propagating in reverse towards the burst   decreases
due to  high density in the ejecta.
This could  explain  the short lifetimes of  some afterglows.
Giannios et al (2008) claim that the early optical afterglow light curves depend critically on the
existence of a reverse shock into a strong magnetized ejecta.
In the first afterglow period the reverse shock plays a main role or  disappears due to
a strong magnetic field
(Giannios et al  2008, Zhang \& Kobayashi 2005, Sari \& Piran 1999, etc).
A reverse shock may also  dominate the emission in
later afterglow phases.
After the reverse shock reaches the back part of the ejecta, there is an initial phase of interaction where shocks
and rarefaction waves cross the ejecta. Then the whole structure relaxes to a self-similar solution.
The X-ray  energy emitted  during the afterglow  is relatively  high (e.g. 2-10 keV at 11 hours after GRB, Piran 2004)
and is most probably emitted during  free expansion.
Optical afterglows disappear after  a few months, therefore  they do not  prevent  to search for GRB host galaxies
(Chen 2018).
Recombination coefficients downstream of
 the shock fronts determine the lifetime of the afterglow.
The recombination lifetime  of hydrogen ions  is t$_{rec}$ $\sim$ 1/(n $\alpha_R$),
where $\alpha{_R}$=4$\times$ 10$^{-13}$ (T/10$^4$)$^{-0.72}$ is the recombination coefficient of H
(Aldrovandi \& P\'{e}quignot 1973).
A very high  density n (adapted to the ejecta) can  reduce t$_{rec}$ to the order of seconds.
In the  low density case, emission features  from the afterglow can be seen even  after  its disappearance.
The afterglow  is characterised by  synchrotron radiation
observable  from several days  to a few months (Chen 2018).
After the first afterglow period
 and an initial stage of  expansion in a dilute ISM, the ejecta will interact
with a  higher density medium in the host galaxy.

\subsection{Host galaxy radiation and collisional ionization sources}

A SB  is generally present in the GRB host galaxy.
The full width at half maximum (FWHM) of the   GRB031203 line profiles
from  Guseva et al (2011, fig. 7) 
indicates a complex  kinetic field  with high and low  velocities.
The  narrow line velocity range is $\sim$  90-130 \kms
while the  broad line  one is $\sim$ 200-330  \kms. 
This suggests that   composite models which
account for the photoionization flux from the SB and  for  shocks, could be  appropriate and could
add some information about  the physical conditions in the host galaxy.
The wind from the  SB stars  affects the host galaxy clouds.
The  gas is  ionized to high levels  by radiation and collisional process.
Recombination times depend on the physical conditions
in the surrounding medium and within the host galaxy. Temperatures can reach  T$>$ 10$^6$K.
The expanding shock   propagates  outwards throughout the  circumstellar medium.
The evolution of the  emitted spectra (line and continuum) observed from the host galaxy is smooth or disturbed
depending  on the  ISM  physical conditions throughout the host.
After a certain time,  the  direct effect of the burst on the emitting gaseous and dusty
clouds  may disappear.

\subsection{Metallicities}

In   GRB hosts,  star formation and  star formation rates (SFR)
 are  at most calculated from the \Ha~ line  and therefore  they  depend on the emitting gas conditions
such as  the density and  the temperature.
Perley et al (2016) investigating whether
the GRB is affecting the SFR independently from other factors,  concluded that
the most "invoked possibility" is a dependence  on metallicity.
GRBs   affect  the host galaxy properties in the surrounding medium.
Therefore, the heavy elements cannot be neglected.
SFR is  proportional to the fractional
 abundance of the different ions in  different ionization levels.
A low metal abundance  "both discourages mass loss from the parent stars and encourage mixing".
Therefore,   Perley et al predicted that GRB should occur in very metal poor environments.
Single star models (MacFadyen \&  Woosley 1999, Maeder et al 2005, Hitschi et al 2005, etc )  imply
that GRBs should occur  mainly in very metal poor environments.
On the other hand, to produce GRB through an energetic jet, a high metallicity is needed,
 because a higher metal
abundance produces stronger stellar winds, greater mass loss and less interior mixing
(Crowther et al 2002, Heger et al 2003, Vink \& de Koter 2005, Hirschi et al 2008).
How much does   the  progenitor's low metallicity affect the metallicity of the host clouds depends
also on  the merging processes.
The metallicity  upper limit  to GRB production  is estimated by Wolf \& Podsidlowski (2007)  to 12+log(O/H)$<$ 8.7.
Chen (2018) reconciled the GRB origin from the death of rapid evolving massive stars
with active forming regions  into the epoch of reionization.

\section{Calculations}

\subsection{Description of the code} 

We use composite models which account consistently for
photoionization and shocks.  The code {\sc suma} (see also Ferland et al 2016)  is adopted.
The main input parameters are those which   are used for the
calculations of the line and continuum fluxes.
They account for photoionization and heating by  primary and secondary  radiation and collisional
process due to shocks.
The input parameters such as  the shock velocity \Vs, the atomic
preshock density \n0 and the preshock
magnetic field \B0 (for  all models \B0=10$^{-4}$Gauss is adopted)
define the hydrodynamical field.
They  are  used in the calculations  of the Rankine-Hugoniot equations
  at the shock front and downstream.
They  are combined in the compression equation (Cox 1972) which is resolved
throughout each slab of the gas
in order to obtain the density profile downstream.
Primary radiation for SB in the GRB host galaxies is approximated by a black-body (bb).
 The input parameters that represent the primary radiation from the SB are the 
 effective temperature  \Ts and the ionization parameter $U$. A pure blackbody radiation 
 referring to \Ts is a poor approximation for a star burst, even adopting a dominant 
 spectral type (see Rigby \& Rieke 2004). However, it is the most suitable because 
 the line ratios that are used to indicate \Ts  also depend on metallicity, 
 electron temperature, density, ionization parameter, morphology of the ionized clouds and, 
 in particular, the hydrodynamical picture.
 For an AGN, the primary radiation is the power-law radiation
flux  from the  active centre $F$  in number of photons cm$^{-2}$ s$^{-1}$ eV$^{-1}$ at the Lyman limit
and  spectral indices  $\alpha_{UV}$=-1.5 and $\alpha_X$=-0.7. The primary radiation source
 does not depend on the host physical condition but it affects the surrounding gas.   This  region  is not considered
as a unique cloud, but as a  sequence of slabs with different thickness calculated automatically
following the temperature gradient. The secondary diffuse radiation is emitted
from the slabs of
gas heated  by the radiation flux reaching the gas and by the shock.
Primary and secondary radiation are calculated by radiation transfer.

In our model the line and continuum emitting  regions throughout the galaxy cover  an ensemble of fragmented clouds.
The geometrical thickness of the clouds is  an input parameter of the code ($D$) which is  calculated
consistently with the physical conditions and element abundances of the emitting gas.
The fractional abundances of the ions are calculated resolving the ionization equations
for each element (H, He, C, N, O, Ne, Mg, Si, S, Ar, Cl, Fe) in each ionization level.
Then, the calculated line ratios, integrated throughout the cloud thickness, are compared with the
observed ones. The calculation process is repeated
changing  the input parameters until the observed data are reproduced by the model results, at maximum
within 10-20 percent
for the strongest line ratios and within 50 percent for the weakest ones.

However,  some parameters regarding the continuum SED, such as the dust-to-gas  ratio $d/g$  and the dust grain radius
\agr ~ are not directly constrained by fitting the line ratios.
Dust grains are heated by the primary radiation and by mutual collision with  atoms.
The intensity of dust reprocessed radiation in the infrared (IR)  depends on $d/g$ and \agr.
In this work we use $d/g$=10$^{-14}$ by number for all the models which corresponds to
4.1 10$^{-4}$ by mass for silicates (Draine \& Lee 1994).
The distribution of the grain size along the cloud starting from an initial radius
is automatically  derived by {\sc suma},
 which calculates sputtering of the grains  in the different zones downstream of the shock.
The sputtering rate depends on the gas temperature, which is $\propto$ V$_s^2$ in the immediate post-shock region.
In the high-velocity case (\Vs$\geq$ 500 \kms) the sputtering rate is so high that the grains
with \agr $\leq$0.1 \mum are rapidly destroyed downstream.
So, only grains with large radius (\agr $\geq$0.1 \mum) will survive.
On the other hand, the grains survive downstream of low-velocity shocks ($<$200\kms).
Graphite grains are more sputtered than silicate grains for T= 10$^6$ K (Draine \& Salpeter 1979).
Small grains (e.g. PAH) survive in the extended galactic regions on scales of  hundred parsecs and lead to the
characteristic features that appear in the SED.
In conclusion, cold dust or cirrus emission results from heating by the  interstellar radiation field,
warm dust is associated with star formation regions and hot dust appears around  AGN (Helou 1986) and in high velocity shock regimes.
Therefore, we will consider relatively large grains, e.g. silicate grains with an initial radius of 0.1 -1.0 \mum.

In the radio range the power-law spectrum of synchrotron radiation
created by the Fermi mechanism at the shock front is seen in most galaxies.
It is calculated by {\sc suma} adopting a  spectral index of -0.75 (Bell 1977).

\subsection{Calculation details}

The calculations initiate at the shock front where the gas is compressed and  adiabatically thermalised, reaching a maximum 
temperature in the immediate post-shock region T$\sim$ 1.5$\times 10^5$ (\Vs/100 \kms)$^2$. T decreases downstream following recombination. 
The cooling rate is calculated in each slab. The downstream region is cut  into a maximum of 300 plane-parallel slabs with 
different geometrical widths calculated automatically, to account for the temperature gradient.
In each slab, compression is calculated by the Rankine-Hugoniot equations for the conservation of mass, momentum and energy 
throughout the shock front. Compression (n/\n0) downstream ranges between 4 (the adiabatic jump) and $\geq$10, depending on \Vs and \B0. 
The stronger the magnetic field, the lower the compression downstream, while a higher shock velocity corresponds to a higher compression.
The ionizing radiation from an external source is characterised by its spectrum and by the flux intensity. 
The flux is calculated at 440 energies from a few eV to keV. Owing to radiative transfer, the spectrum changes throughout 
the downstream slabs, each of them contributing to the optical depth. In addition to the radiation from the primary source, 
the effect of the diffuse radiation created by the gas line and continuum emission is also taken into account, using 240 energies 
to calculate the spectrum.  For each slab of gas, the fractional abundance of the ions of each chemical element is 
obtained by solving the ionization equations  X$_{i+1}$/X$_i$=X${_I}$+X$_{II}$+X$_{III}$ (Contini \& Shaviv 1980), where
X$_I$ represents the contribution to ionization from the primary source (stars or  AGN) (Williams 1973), X$_{II}$
represents the contribution of collisional ionization  (Cox \& Tucker 1969) and X$_{III}$ is the contribution
of secondary radiation (Williams 1967). The parameters are described by Contini \& Aldrovandi (1983 and references 
therein) and Viegas-Aldrovandi \& Contini (1989 and references therein).

These equations account for the ionization mechanisms (photoionization by the primary and diffuse radiation, and collisional
 ionization) and recombination mechanisms (radiative, dielectronic recombinations), as well as charge transfer effects. 
The ionization equations are coupled to the energy equation if collision processes dominate, and to the thermal balance if 
radiative processes dominate. The latter balances the heating of the gas due to the primary and diffuse radiations reaching the 
slab with the cooling due to recombinations and collisional excitation of the ions followed by line emission, dust collisional 
ionization and thermal bremsstrahlung. The coupled equations are solved for each slab, providing the physical conditions necessary 
for calculating the slab optical depth, as well as its line and continuum emissions. 
The slab contributions are integrated throughout the cloud.  
In particular, the absolute line fluxes corresponding to the ionization level i of element K are calculated 
by the term nK(i), which represents the density of the ion i. 
We consider that nK(i) = X(i) [K/H] n$_H$, where X(i) is the fractional abundance of the ion i 
calculated by the ionization equations, [K/H] is the relative abundance of the element K to H and n$_H$ 
is the density of H (in number cm$^{-3}$). In models including shock, n$_H$ is calculated by the compression equation (Cox 1972) 
in each slab downstream. Accordingly, the abundances of the elements are given relative to H as input parameters.

Dust grains are coupled to the gas across the shock front by the magnetic field (Viegas \& Contini 1994). 
They are heated by radiation from the AGN and collisionally by the gas to a maximum temperature, which 
is a function of the shock velocity, of the chemical composition and of the radius of the grains, 
up to the evaporation temperature (T(dust) $\geq$ 1500 K). The grain radius distribution downstream 
is determined by sputtering, which depends on the shock velocity and on the density. Throughout shock fronts 
and downstream, the grains might be destroyed by sputtering.

Summarizing, the code starts by adopting an initial \Te (10$^4$ K) and the input parameters for the first slab. 
It then calculates the density from the compression equation, the fractional abundances of the ions 
from each level for each element, 
line emission, free-free emission and free-bound emission. It re-calculates \Te by thermal balancing or the enthalpy equation, 
and calculates the optical depth of the slab and the primary and secondary fluxes. 
Finally, it adopts the parameters found in slab i 
as initial conditions for slab i + 1. 
Line and continuum intensities are integrated  accounting for all the  slabs.  The number of lines  calculated by 
{\sc suma}  is  $>$ 200 for each model.
 They are calculated  at the gaseous nebula that emits the spectrum, while
 the data are observed at Earth. Therefore they diverge by a factor (r$^2$/d$^2$) that depends on the distance of the 
nebula from the  radiation centre (r), 
and on the distance (d) of the galaxy to Earth. We then calculate the line ratios to a specific 
line (in the present case \Hb, 
which is a strong line), and compare them with the observed line ratios.

 On this basis we calculate a grid of models. The  set of  models   (e.g. Table 1)
which best reproduce the line ratios is selected. We  obtain the final model by cross-checking
the fit of the calculated continuum  SED  to the observed one.

\subsection{Modelling the continuum SED}

   The  models constrained by the  line spectra
   give a hint about the relative importance of the different
ionization and heating mechanisms which  are recognised throughout the continuum SED
in each of the objects. 

The gas ionized by the  SB (or AGN) radiation flux  emits continuum radiation (as well as the line fluxes)
from  radio to  X-ray.
The continuum accounts for
  free-free and free-bound radiation (hereafter  addressed
to as bremsstrahlung).
The bremsstrahlung  at $\nu$$<$ 10$^{14}$ Hz has a similar slope in
all the diagrams.
In fact, the
bremsstrahlung continuum, emitted by free electrons accelerated in
Coulomb collisions with positive ions
(mostly H$^+$, He$^+$ and He$^{++}$) in nebulae of charge Z has an
emission coefficient (Osterbrock 1974):

J$_{\nu}$ $\propto$\Ne N$_+$Z$^2$($\pi h\nu$/3kT)$^{1/2}$ e$^{-(h\nu/kT)}$    (1)

\noindent
The  photoionization radiation flux can heat the gas to T$\sim$ 2-4$\times$ 10$^4$ K, while the
gas is heated  collisionally by
the shock to a maximum of T$\propto$ (\Vs)$^2$, where
\Vs is the shock velocity.
The cooling rate downstream  depends on  \Ne N$_+$ (N$_+$ is the proton density).
The trend of the bremsstrahlung as function of $\nu$  follows the
interplay between T  and $\nu$.
High temperatures of the emitting gas  determine the  bremsstrahlung maximum  at high $\nu$.
At T$\sim$1-4$\times$10$^4$ K the exponential term is significant
at frequencies between  10$^{14}$ and 10$^{15}$ Hz.
The  temperatures  are calculated by thermal balancing between the heating rates which depend on the
photoionizing flux and  the cooling rates by free-free, free-bound and line emission.
Therefore,  the radiation effect is seen mainly in this frequency  range.
In the radio range, the exponent  in eq (1)
tends to 0 and  the continuum is $\propto \nu^{1/2}$.
So the SEDs in all the  diagrams  of all  galaxy types have  similar trends  at
relatively low frequencies and  the dust reprocessed radiation bump in the IR is clearly recognizable.

In conclusions,
i) the black body radiation corresponding directly to the temperature
dominating in the star-burst  is seldom observed  in the
UV, because  absorption  is very strong in this frequency range
due  to  strong line formation.
ii) The shock effect throughout the SED can be recognized from  the maximum frequency and
intensity of the dust reprocessed radiation peak
in the infrared and of the bremsstrahlung   at high frequencies.
iii) 
The gas ionized by the  SB (or AGN) radiation flux  emits  bremsstrahlung
from  radio to  X-ray.
The black body emission from the background old star  population with
T$_{bb}$$\sim$ 3000-8000 K  generally emerges
over  the bremsstrahlung  throughout the   SED in the near-IR(NIR) - optical range.
iv) In the radio range synchrotron radiation created by the Fermi
mechanism  is recognized by its spectral index.
   Thermal bremsstrahlung in the radio range has a steeper trend which becomes even steeper
by self-absorption at low $\nu$.
In the far-IR only comparison with the observation data  indicates the source of the continuum
radiation flux, because  thermal bremsstrahlung, synchrotron radio and cold dust reradiation may be
blended.

\section{Modelling GRB031203 line spectra}

\subsection{Spectral  characteristics}

In the following  we  present  the results obtained by the   detailed modelling of the LGRB031203 host galaxy 
spectra observed  by Prochaska et al (2004)
on the Magellan/Baade 6.5m telescope and Inamori Magellan Areal Camera and Spectrograph (IMACS), at different epochs  (2003-2004) by Margutti et al (2007) on the 
ESO-Very Large Talescope (VLT) 
focal reducer/low-resolution spectrograph (FORS) and by Guseva et al (2011) by the VLT X-shooter. 
Both narrow and broad lines  were observed at later times   by Guseva et al, in March 2009.
Watson et al (2011) presented the first spectroscopic study  of the  LGRB031203 host galaxy in the mid-IR (5-40 \mum)
by VLT X-shooter on 2009 March 17
and Spitzer photometric observations from the UV -mid-IR (0.35-40 \mum) with detections and upper limits at far-IR,
submillimeter and radio wavelengths. They were carried out on 2005 May 28.
Observation epochs and  \Hb observed fluxes are summarised in Table 1.

The sketch in Fig. 2 (right  panel)  represents  the  clouds  emitting the
 narrow lines (models mod0-mod6 in Tables 2-4), while  the  left panel   explains the  broad lines
emission (model mod7 in Tables 3 and 4). 
Both strong and weak shock fronts are  reached by the photoionization flux from the SB.
In the case  of a SB, the clouds (representing the debris from the SB)  propagate outwards
with velocity V$_{debris}$$\sim$100 \kms from the radiation source i.e. from the SB.
The SB debris  collide with the ISM clouds which are nearly standing
with velocities V$_{ISM}$$\sim$30\kms. A head-on shock follows with \Vs=V$_{debris}$+V$_{ISM}$$\sim$100-130\kms.
Moreover, a relatively high velocity wind (V$_{wind}$$\sim$400 \kms) originating from the SB stars reaches the  expanding
debris  yielding a head-on-back shock with \Vs=V$_{wind}$-V$_{debris}$$\sim$ 300 \kms.
A high velocity wind at earlier times ($\sim$ 1000 \kms)
could have easily  destroyed the gaseous clouds by collisions.
It was shown by Schiano (1985, 1986) that clouds with preshock densities $<$ 100 \cm3 are swept from the region
of free  flowing winds (Contini \& Viegas-Aldrovandi 1990).
Regarding line profiles in wind regimes, although in a different contest,
 Vogel \& Nussbaumer (1994) noticed that the luminosity of
the hot component of the symbiotic star AG Peg decreased from 1600\Lsol in 1978 to 500\Lsol in 1990 and furthermore to 400\Lsol in 1993.
NV and HeII lines  exhibited a pure wind profile  on top of
which a nebular contribution appeared in 1981 and 1986, respectively. These profiles were
recognised by Penston \& Allen (1985) as a convolution of a nebular component and a broad line component,
which they attributed to a wind lost by the hot companion with a  velocity of $\sim$ 900 \kms.
In all the observations the wind velocity retained this value, although the strength 
gradually decreased.
This decline  was  followed by qualitative changes in the profiles of
the emission features.

\begin{table*}
\centering
\caption{Observation details (2003-2009)}
\begin{tabular}{lccccccccccccc} \hline  \hline
\         &  obs$^0$        &  obs$^1$          &  obs$^2$       &  obs$^3$           & obs$^4$          &   obs$^5$   &  obs$^6$   \\ 
\  day    &2003 Dec 6.3     &2003 Dec 20        &2003 Dec 30     & 2004 Mar 2         &2004 Sep 2        &  2009 Mar 17&2009 Mar 17   \\
\ \Hb$^7$ & 51.8$\pm$0.8    & 2142.39$\pm$326.70&  2135.91$\pm$10& 1932.24$\pm$571.15 &1949.33$\pm$263.49 &3.85$\pm$0.04& -\\ 
\ Telescope & Magellan/Baade&ESO-VLT            &ESO-VLT       & ESO-VLT              &ESO-VLT            &VLT             & VLT       \\ 
\           & IMACS         &        FORS2      &        FORS1 &         FORS1        &       FORS1      & X-shooter      & X-shooter \\ \hline
\end{tabular}  

$^0$ Prochaska et al (2004);
$^1$ Margutti et al (2007); 
$^2$ Margutti et al (2007);
$^3$ Margutti et al (2007);
$^4$ Margutti et al (2007);
$^5$ Guseva et al (2011);
$^6$ Watson et al (2011);
$^7$ in 10$^{-16}$ \erg observed at Earth

\end{table*}

\begin{table*}
\centering
\caption{Extinction corrected line ratios to \Hb at different epochs from Prochaska et al (2004) and Margutti et al (2007)}
\begin{tabular}{lcccccccccccccccccc} \hline  \hline
\              & obs$^0$  &  mod0   & obs$^1$  &  mod1 & obs$^2$& mod2  & obs$^3$&mod3   & obs$^4$&mod4  \\ \hline
\ [OII]3728.8  &1.06      & 1.06    &1.20      & 1.26   &1.33   &1.33   & 1.20   &1.23   & 1.11   &1.10\\
\ [NeIII]3868+ &0.83      & 0.90   &0.98      & 1.00   &0.88   &0.97   & 0.84   &1.     & 0.83   &1.00\\
\ \Hg          &0.49      & 0.46   &0.46      & 0.46   &0.45   &0.46   & 0.43   &0.46   & 0.45   &0.46\\
\ [OIII]4363   &0.11      & 0.08   &0.08      & 0.08   &0.08   &0.08   & 0.07   &0.08   & 0.08   &0.08 \\
\ HeI 4471.5   &0.05      & 0.05   &0.03      &0.05   & -      &  -    & -      &-      & -       &-  \\
\ \Hb          & 1        &   1    &1         & 1      &1      &  1    & 1      & 1     & 1      &1   \\
\ [OIII]5007+  &8.46      & 8.40   &8.72      & 8.50   &8.52   & 8.69  &8.61    &8.55   & 8.53   &8.50\\
\ HeI 5876     &0.12     & 0.14    &0.11      & 0.14   &0.12   & 0.14  &0.12    &0.14   & 0.14   &0.14 \\
\ [OI] 6300+   &0.02      & 0.01   &0.04      & 0.02   &0.03   & 0.016 &0.04    &0.01   & 0.03   &0.01\\
\ [SIII]6312   &0.02      & 0.02   &0.02      & 0.02   &0.02   & 0.02  &0.02    &0.02   & 0.01   &0.02 \\
\ [NII]6548+   &0.21      & 0.21   &0.21      & 0.25   &0.19   & 0.20  &0.18    &0.20   & 0.19   &0.18 \\
\ \Ha          &2.82      & 2.90   &2.83      & 2.9    &2.78   & 2.91  &2.79    &2.89   & 2.96   &2.89 \\
\ [SII]6717    &0.08      & 0.08   &0.09      & 0.08   &0.10   & 0.08  &0.10    &0.1    & 0.10   &0.10 \\
\ [SII]6731    &0.07      & 0.09   &0.07      & 0.08   &0.07   & 0.08  &0.08    &0.08   & 0.08   &0.08 \\
\ [AIII]7136   &0.07      & 0.06   &0.07      & 0.10   &0.07   & 0.07  &0.07    &0.1    & 0.06   &0.10 \\
\ [OII]7320+   &-         &-       &0.03      & 0.04   & 0.04  &0.03    &0.03   & 0.02   &0.02   & - \\ \hline
\end{tabular}

obs$^0$ - obs$^4$  observations are described in Table 1;

\end{table*}

\begin{table*}
\centering
\caption{Modelling extinction corrected line ratios to \Hb  from Guseva et al (2011)}
\begin{tabular}{lccccccccccccc} \hline  \hline
\              &obs$^5$  & mod5  & obs$^6$& mod6 & obs$^7$   &mod7    \\ \hline
\ [OII]3728.8  &0.86     &086    & 0.98   &0.99  &-          &1.03\\
\ [NeIII]3868+ &0.38     &0.40   & -       &-    &-          &0.49\\
\ \Hg          & 0.50    &0.46   & -       &-    &0.48       &0.46\\
\ [OIII]4363   & 0.09    &0.10   &  -      &-    &-          &0.07\\
\ HeI 4471.5   & 0.04    &0.05    &  -      &-    &-         &0.05 \\
\ [ArIV]4711   &0.016    &0.02   &  ...    &-     &-         &0.1\\
\ \Hb          & 1       & 1     & 1       &1     &1         & 1 \\
\ [OIII]5007+  & 9.78    &9.20   & 5.89    &5.60  &10.20     &10.30\\
\ HeI 5876     & 0.12    &0.13   &-       &-      &0.10      &0.14\\
\ [OI] 6300    & 0.03    &0.01  & 0.03    &0.01   &   -      &0.01 \\
\ [SIII]6312   & 0.02    &0.02   & -       & 0.02 &    -     &0.01  \\
\ [NII]6548+   & 0.18    &0.17   & 0.13    & 0.17 &    -     &0.17\\
\ \Ha          & 2.85    &2.92   & -       & 2.96 &    -     &2.90\\
\ [SII]6717    & 0.09    &0.08   & 0.18    & 0.16 & 0.20     & 0.70 \\
\ [SII]6731    & 0.07    &0.08   &  -      &  -   & 0.20     &0.10 \\
\ [AIII]7136   & 0.06    &0.05   & -       &  -   &0.08      &0.06 \\
\ [OII]7320+   &0.03     &0.03   &  -      & -    & -        &0.05\\
\ [SIII]9069   &0.25     &0.26   &  -      &-     &0.45      &1.00\\
\ [SIV]10.61   &0.47     &0.60   &         &   -  &-         &0.50\\
\ [NeII]12.8  &0.04     &0.04   &         &   -   &-         &0.04  \\
\ [NeIII]15.56 &0.59     &0.53   &    -   &   -   &-         & 0.80      \\
\ [SIII]18.71  &0.29     &0.27   &    -   &   -   &-         &0.36 \\
\ \Hb$^8$      &         &0.07   &        &0.06   &-         &0.25 \\  \hline
\end{tabular}

obs$^5$  observations by Guseva et al and obs$^6$ reported by Watson et al are described in Table 1;
obs$^7$  the broad line ratios presented by Guseva et al.; $^8$ \Hb absolute flux calculated at the nebula in \erg.

\end{table*}

\begin{table*}
\centering
\caption{Model results}
\begin{tabular}{lccccccccccccc} \hline  \hline
\                    & mod0   & mod1        &mod2        &mod3      &mod4    &mod5    & mod6 &mod7 & GIFH$^7$\\ \hline
\ \Hb$^5$            & 0.07   &  0.04       & 0.043      &0.02      &0.02    &0.07    &0.06  &0.25 &   -\\
\  \Vs (\kms)        &130     & 120         & 120        &100       & 90     &110     &110   & 300&   -\\
\ \n0  (\n0)         &100     &  90         & 90         & 60       &50      &90      &90    & 100& -  \\
\ $D$  (pc)          &2.93     & 0.44        &2.93        &1.03      &1.71    &4.30     &4.00    & 1.40  &  -\\
\ \Ts  (10$^4$K)     &7.40    & 7.40        & 7.40       &7.20      & 7.20   &6.50    &5.50  &6.50  &   -\\
\ $U$ -              & 0.06   & 0.04        & 0.04       &0.03      &0.03   &0.08     &0.02  &0.16&- \\
\ He/H               &0.1     &0.1          &0.1         &0.1       &0.1    &0.1      &0.1  &0.1& 0.1 \\
\ N/H$^6$            &0.14    &0.14         &0.14        &0.12      &0.12   &0.14     &0.14   &0.14  & 0.12   \\
\ O/H$^6$            &2.0     & 2.0         & 2.0         &2.0       &2.0    &2.0      &2.0   &3.0  &1.6    \\
\ Ne/H$^6$           &0.4     & 0.6         & 0.4        &0.6       &0.6    &0.3      &0.4  &0.3  &0.3      \\
\ S/H$^6$            &0.05    & 0.04        &0.05        &0.06      &0.05   &0.07     &0.08 &0.08 &0.04       \\
\ Ar/H$^6$           & 0.007  & 0.060       &0.006        &0.060     & 0.007  &0.006  &0.006&0.006&0.004       \\ \hline
\end{tabular}

$^5$ \Hb absolute flux calculated at the nebula in \erg ;
$^6$ in 10$^{-4}$ units; $^7$ model results for Guseva et al (2011);

\end{table*}

\subsection{Line ratios from Prochaska et al (2004)  and Margutti et al (2007) observations}

Margutti et al (2007) found that  GRB031203 host  is metal poor with 12+log(O/H)=8.12.
The GRB luminosity is low and the decay rate unusually slow, while
underluminous GRB with fast decay rates at early times (t$\leq$0.5 days) have been suggested by Bloom et al (2003). 
In Tables 2 and 3 we compare model results with  reddening-corrected line ratios in the next column. 
The  observations by VLT  
presented by Prochaska et al  (2004) in 2003  are followed  in Table 2 by  those  of Margutti et al (2007, their table 4) 
at different epochs in 2003-2004.
The fit is quite satisfying   considering that the calculated line ratios are
within the errors (20 percent for strong lines and 50 percent for weak lines). 
The observed line ratios  constrain the models
because \Ha~  and \Hb are both observed in the spectra and oxygen  is present  by three  ionization level lines. 
 [OIII] 5007+/[OIII]4363 line ratios  are also available. It is known that they  are
low when the gas is heated
to relatively high temperatures. This   may occur  when   shocks  are accounted for and photoionization
is weak. Table 2 shows, on the contrary, that [OIII]5007/[OIII]4363 are relatively high ($>$80)
in all the spectra,  indicating that photoionization dominates.
Sulphur  appears  by the S$^+$ and S$^{++}$ ions  in  the [SII]6717, 6731 doublet  and in  the [SIII] 6312 line,
respectively. 
Although the [SIII]6312 line   is generally weak and blended with the [OI]6300,6360 doublet,
  this line has been  measured in the specific case of GRB031203, constraining the model.

The results of modelling  are presented in Table 4. The input parameter trends follow 
 the small changes of
the observed line ratios with time  (Table 4).  Those are within the observed errors  
and   indicate that the photoionizing source of the host galaxy gas did not significantly vary
during the  observation period. 
A SB is generally adopted to explain the photoionizing flux in GRB hosts
by the observer community. 
Our calculations confirm (see Guseva et al 2011)  that  a SB is  adapted to explain 
GRB031203  host spectra.
The ionization parameter  decreases  with   the emitting gas distance from the hot source and/or because
the radiation flux is prevented to reach the cloud by some obstructing matter. 
Accordingly, Table 4 shows a small decreasing trend of $U$  with time from December 2003 to September 2004, 
as the shock propagates throughout the host ISM.
The SB effective temperature decreases  after September 2004.
With regards to the shock parameters, we have found that
shock velocities and preshock densities  slightly decrease  from December 2003 towards a minimum in 
September 2004.
The geometrical thickness of the clouds, on the other hand, changes  randomly due to fragmentation  created
by turbulence  at the  shock fronts.

All the  line ratios observed in a single-time spectrum  are  reproduced by models  implying
that the radiation flux from the SB   and the shock are coupled.
Our models  show [SII]6717/[SII]6731 ratios $<$ 1 in some spectra, whereas the observed ones are $>$1, 
indicating very low densities of the emitting gas. Also the [OI]/\Hb line ratios are underpredicted.
We explain  these results by the contribution of a large amount of ISM gas where the  [OI] lines are
strong and the densities are low.

In Figs. 3 and 4 the profiles of the electron temperature and electron density (top panel) and of the fractional abundances 
of the  most significant ions (bottom panel)  throughout a cloud are shown.
Fig. 3  presents the case of the shock created by collision of the debris with the ISM clouds
which  contributes to reproduce most of the observed spectra (mod0-mod6). The
gas  shows a maximum downstream temperature of $\sim$ 1.5 10$^5$K near the shock front 
(T$\sim$ 1.5$\times 10^5$ (\Vs/100 \kms)$^2$). 
Therefore,
the O$^{++}$ ion  dominates  throughout a large region of the cloud (Fig. 3).
The density downstream which depends on the shock velocity  is relatively low ($<$10$^3$\cm3).
The gas downstream cools down  following slow recombination because the density is low.
The large region  of gas at T$\sim$10$^4$K explains why models accounting on pure photoionization
reproduce the lines ratios as well as those accounting for the coupled effect of shock and photoionization.
Fig. 3 shows that the gas is not fully  recombined at a distance from the shock-front $\geq$1 pc 
due to  the low compression downstream which depends on \Vs~ and \n0.
It was found by modelling the line ratios that the emitting clouds   are matter-bounded.

\begin{figure}
\centering
\includegraphics[width=8.8cm]{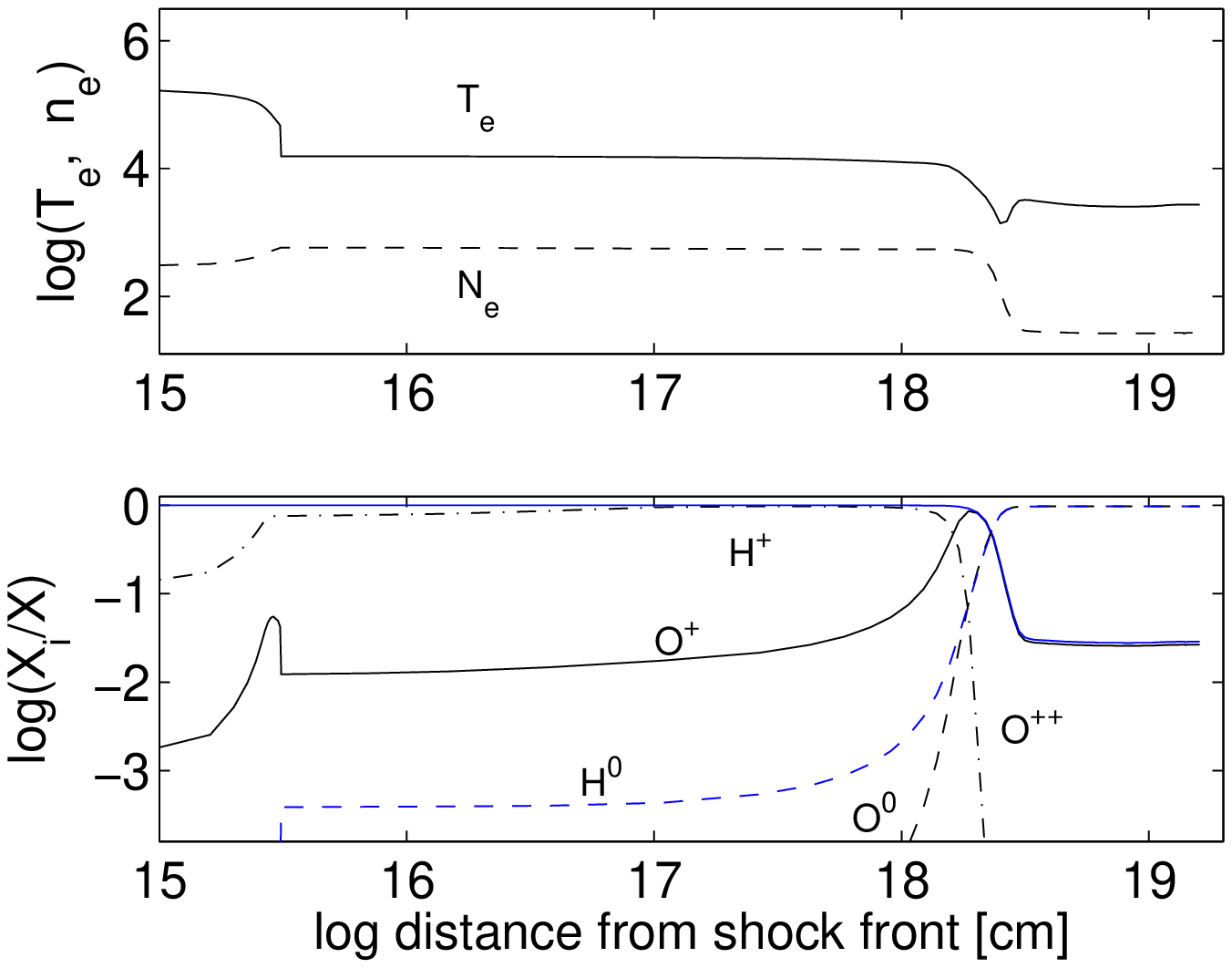}
\caption{The shock front is on the left. Top
panel: the distribution on \Ne and \Te throughout the
cloud calculated by model mod5. Bottom panel: the distribution of the
 H$^0$/H, H$^+$/H, O$^0$/O, O$^+$/O and O$^{++}$/O fractional abundance.
}
\centering
\includegraphics[width=8.8cm]{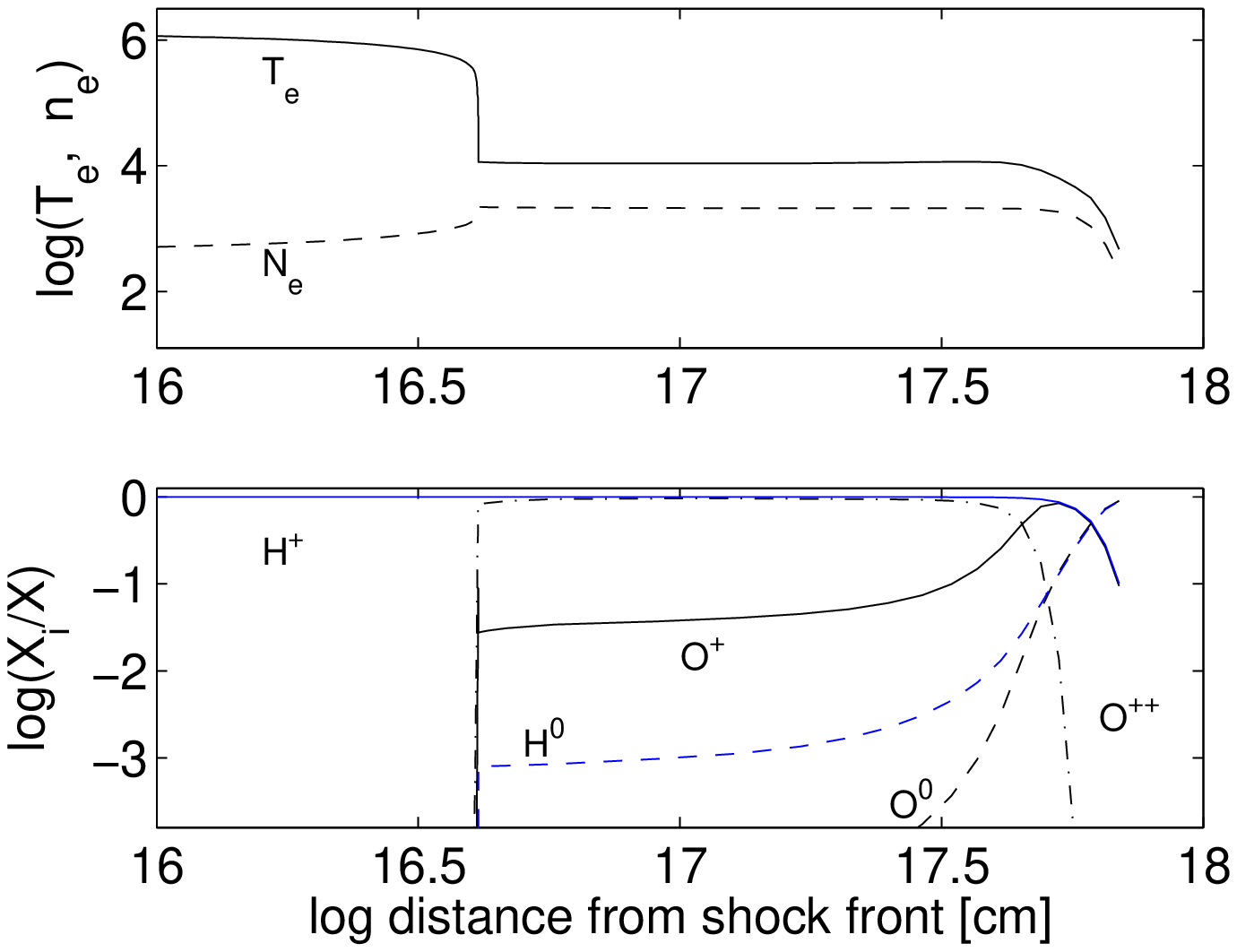}
\caption{The same as  in Fig. 3 for model mod7.
}
\end{figure}

\subsection{Line ratios from Guseva et al (2011) and Watson et al (2011) observations}

Guseva et al presented  VLT/X-shooter observations performed  on March 2009.
The same observations were discussed by Watson et al (2011) who added Spitzer spectroscopy
of strong forbidden lines in the mid-IR in their table 3.
We compare model results to the reddening-corrected observed line ratios in Table 3.

The line ratios were  previously modelled by Guseva et al  adopting the  code CLOUDY. They obtained very good results
for all of them, 
therefore, they concluded that photoionization is the only source of the emitted lines.
The fit is   also good adopting  a composite  model (photoionization + shock). 
In fact, most of the lines
observed are emitted from ions in the intermediate ionization levels II and III.  
 Those lines are mainly produced  by the photoionizing flux which heats the
gas to $\leq$2-3$\times10^4$K. Shocks can heat the gas to higher temperatures depending on the shock velocity,
therefore shocks are mainly revaled by strong high ionization level lines and neutral lines (see Fig. 4).
Moreover, a composite model can fit the observed SED from radio to X-ray, because the  bremsstrahlung maximum
reaches relatively high frequencies depending on the shock velocity. The mutual heating and cooling between
gas and dust grains lead to the bump in the IR calculated consistently with the line emission. Also,
synchrotron radiation created by the Fermi mechanism at the shock front  is often observed in the radio range.

The intensities of H recombination lines were calculated by Guseva et al by an
electron temperature \Te=12500K, and electron density \Ne=100 \cm3 and case B  which indicates
large optical depths.
An oxygen metallicity  12+log(O/H)=8.20 was  evaluated by the \Te method, in agreement with previous 
results (Prochaska et al 2004, Hammer et al 2006, etc). 
The element relative abundances calculated by Guseva et al (2011) are reported in the last column (GIFH) of 
Table 4 for comparison.
The N/O abundance ratio results 0.075. Oxygen metallicity   is  lower than that calculated by {\sc suma}
 because the line  ratios presented in Tables 2 and 3 are calculated integrating on the temperature 
decreasing gradient downstream. 
Chen (2018)   suggested that satellite galaxies or tidal tails  contain relatively more pristine gas
around otherwise chemically evolved star-forming galaxies (Thilker et al. 2009).
This is  true  also  for  N/H.  

However, the dynamical phenomena which follow  the SB  require the presence of shocks
throughout the ISM host. 
Analysing the  profiles of the strongest lines, Guseva et al disclosed that relatively high velocities
($\leq$400 \kms) contributed to the FWHM of the  line profiles. They  are  seen as broad sockets in 
Guseva et al (2011) their fig. 7.
The observed relatively large velocity  distribution may be due to fragmentation following turbulence in the  host ISM. 
Interestingly, complex gas kinematics were observed from the absorption lines in the GRB host environments of faint galaxies
at z$\geq$1.5 (Chen 2018) with a   significant spread of velocities (e.g. Prochaska et al 2008).
Chen (2018) claims that  the relatively high velocity (200-400 \kms)    distribution in the  
foreground gas  of the afterglows
is   "starburst driven outflows", most likely  in the host galaxy. This scenario could  produce shocks   
by collision with the ISM  debris and clouds.
Therefore we adopted composite models
which account for shock and photoionization to explain the  spectra.
A black-body radiation is used to represent the flux from  the SB.
We have modelled Guseva et al line ratios by the same type  of models (mod5 and mod6)
that were used for Margutti et al. spectra (Table 2).
 The broad line ratios are calculated by  model mod7 with a velocity \Vs=300 \kms,
which  is  similar to the velocities  yielding  the  sockets in the line profiles.

\begin{figure*}
\centering
\includegraphics[width=8.8cm]{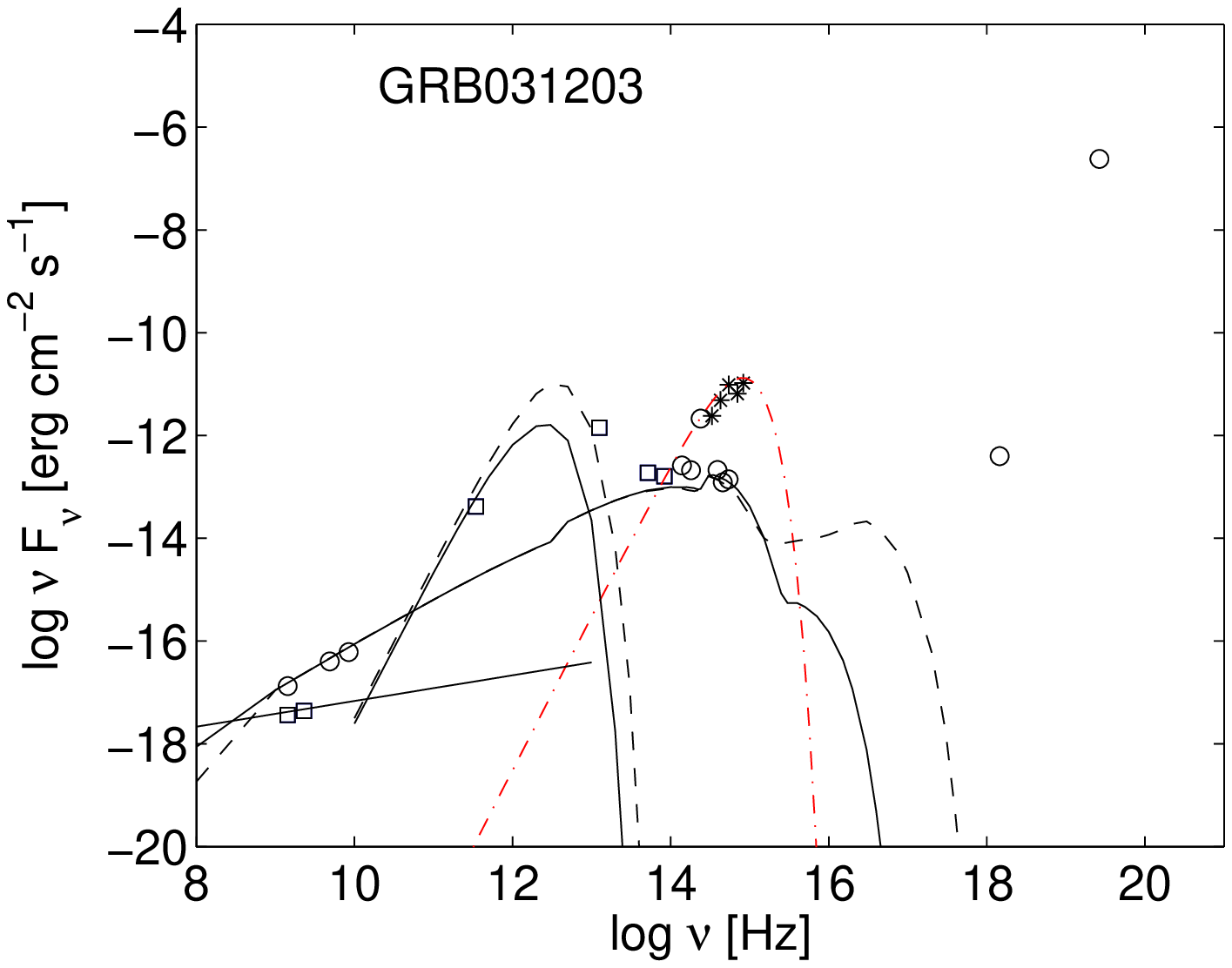}
\includegraphics[width=8.8cm]{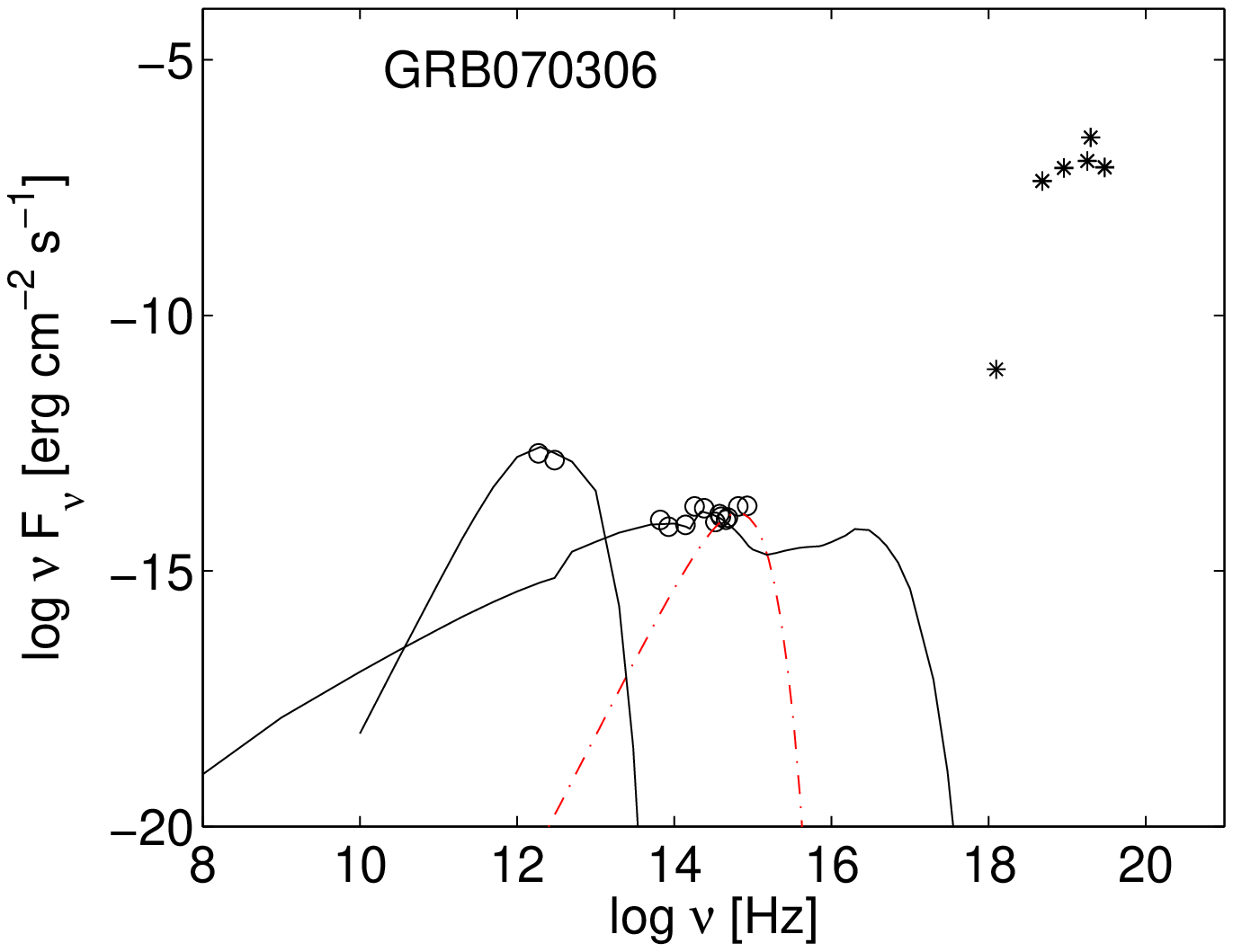}
\caption{Left: SED of GRB031203.
Black open circles in the X-ray: data from Firmani et al (2006) and Gendre et al (2006);
black squares : data  (VLA-Spitzer) from Watson et al;  black asterisks:
data from Margutti et al and Mazzali et al (2006);
	 red dash-dotted line: bb flux calculated with  T=10$^4$K representing the underlying old star population.
Black solid lines: the bremsstrahlung and dust reradiation  calculated by model mod1 (Table 4);
black dashed lines:
the bremsstrahlung and dust reradiation calculated by model mod7 (Table 4);
	black solid line in the radio:  synchrotron radiation  by the Fermi mechanism.
Right: SED of LGRB070306 adapted from Contini (2018a).
}
\end{figure*}

We assume that a relatively strong wind collides with the  clouds on their internal edge
 facing the SB (Fig. 2, left panel).  The shock front on the opposite (outer) edge is
not significant because the cloud propagates in the outer ISM where the electron density is low.
The shock front facing the SB  is   reached
by the photoionization radiation (Fig. 2, left panel). The temperature ($\propto$ \Vs$^2$)  
in the  downstream region
close to the shock-front is relatively high, with a maximum T= 1.35$\times$10$^6$K.
We  report the calculated line ratios (model mod7) in Table 3 and the results of modelling in Table 4. 
The observed [SIII]9069/\Hb line ratio is overestimated by  model mod7 by a factor $\sim$2. 
The other mid-IR lines are satisfactorily reproduced.

The results calculated by the code {\sc suma}  depend on
the stratification of the temperature and the density downstream which lead to the consistent fit of  
most of the line ratios. Fig. 4  shows the \Te and \Ne distribution  for model mod7 which fits 
Guseva et al broad line observations. 
The gas  cools down slowly near the shock front because recombination coefficients are small for T$\geq$10$^5$K.
Then, following  the strong cooling rate due to UV and optical  line emission, the temperature drops to $\leq$10$^4$K. 
The secondary diffuse radiation  maintains the gas at $\geq$ 10$^4$K  throughout a large region.
The gas   recombines at a distance from the shock-front $\leq$1 pc. 
The model which best reproduces the data  is radiation-bounded (Table 4).
In the bottom diagram of Fig. 4 the distribution of the  fractional
abundance of  H$^0$/H, H$^+$/H, O$^0$/O, O$^+$/O and O$^{++}$/O ions  is shown.

\section{Continuum SED}

Spitzer photometric data for GRB031203 were  observed by Watson et al (2011) by IRAC.
Radio continuum observations were obtained on January 2008 using ATCA (Watson et al 2011, table 2).
In Fig. 5 (left panel) we present the modelling of GRB031203 continuum SED  on the basis of Watson et al data.
To reproduce the data, we use  the bremsstrahlung calculated by the models that were
selected by fitting the line ratios.
The continua  calculated by models mod1  and mod7 are  selected because the former represents  
emission at  earlier times 
 and  the latter  represents the  broad line  component which appears  later
 in the line profiles. 
 For each model two curves  are shown in the figure, one is  gas bremsstrahlung which covers the frequency range 
from radio to soft X-rays, the other in the IR
is reprocessed radiation by dust consistently calculated. 
 The bremsstrahlung calculated by the models  shows  similar trends in the  radio-optical range.
It  reproduces  the GRB031203 data in the radio  and  defines the bottom limit to
the IR  data which are generally nested within  a black-body  curve. 
This  bump is explained by the  flux emitted from the  underlying old star population. 
For GRB031203  the old star best fitting temperature is  10$^4$ K. 
The continuum calculated by model mod7 reaches higher frequencies  than   mod1 because the high velocity shock heats 
the gas to higher temperatures.
Synchrotron   radiation  in the radio range, created by the Fermi mechanism at the shock front, is most likely
fitted by mod7. Thermal bremsstrahlung at $\nu$ $<$ 1GHz shows self absorption by  high density.
In the right panel of Fig. 5 we report the SED of LGRB070306 
(Contini 2018a, fig.1) for comparison. 
The underlying old star contribution hardly emerges from the  LGRB070306 bremsstrahlung.
The LGRB031203  two data  in the X-ray range (Firmani et al 2006, Gendre et al 2006) cannot be reproduced 
by the models presented in Table 4 because they are emitted from the afterglow.
In the  GRB070306 SED  (Fig. 5 right panel) and for other LGRB reported by Contini (2018a),  various data 
in the X-ray  appear  at higher frequencies. Only a few of them  are   observed in
GRB031203. In fact, they represent the afterglow  which is weak in  GRB031203.

\section{Concluding remarks}

The spectra observed from the long GRB031203 (z=0.1055) host galaxy are emitted from clouds moving 
with different velocities throughout the ISM. 
It was  explained by Chen (2018)  that  the distribution of velocities depends on
 SB  driven outflows. The lines  observed by Guseva et al in March 2009  show
FWHM profiles of $\sim$ 100 \kms, similar to those observed in 2003-2004 by Margutti et al, but  with a
FWHM $\sim$ 300 \kms component.
We have reproduced the observed  line ratios at different times by two main models.  We   suggest that the 
spectra observed at early times by Margutti et al are emitted downstream of  head-on shock  created 
by  collision of the SB ejected debris   with  nearly standing ISM clouds.  
The shock velocity is  mainly determined by the SB debris.
The spectra corresponding to FWHM of $\sim$ 300 \kms are emitted downstream of  the head-on-back shock 
 created by collision of the fast wind from  SB stars  with the ejected debris. 

The results obtained by the detailed modelling of LGRB031203 host  line ratios observed
at different times are presented in Table 4. 
The models  account for the coupled effects of shocks and photoionization from the SB.
The  line ratios are enough to constrain the models
because \Ha, \Hb  and oxygen  in  three different ionization levels appear in the spectra. 
The [OIII] 5007+/[OIII]4363 line ratios  are  available. 
Sulphur  is   present  by the S$^+$ and S$^{++}$ ions as the [SII]6717, 6731 doublet  and the [SIII] 6312 line. 
Although the [SIII]6312 line   is generally weak and blended with the [OI]6300,6363 doublet,
  this line  is measured in the specific case of GRB031203.
The results show that the main physical parameters (\Vs, \n0 and the ionization parameter) show a small 
decreasing trend from  December 2003  to a minimum in September 2004.
 The SB effective  temperature decreases from September 2004 to March 2009. The element abundances relative to H
 are calculated at each epoch.
S/H slightly decreases after September 2004.
 The O/H ratio is found higher   in the  high velocity clouds than in the low  ones by a factor of 1.5, but it is
still lower than solar by a factor of $\sim$2.
The comparison  of the results obtained by the detailed modelling of the spectra with the results obtained  
using  the \Te method by the observers
 shows that   oxygen  is less depleted in GRB0031203 host than  predicted by previous analysis. 
Prochaska et al (2004) and Guseva et al (2011) found log(O/H)+12=8.2, Margutti et al (2008)  estimated 8.12.
We found 8.3. 
 In Fig. 6 we report N/O results for LGRB hosts, SB, low-luminosity nearby galaxies and HII regions in local galaxies (see Table 5)
as a function of z. It was found that N/O in  LGRB hosts  follows a decreasing trend with decreasing z,
while N/O  for supernovae (SN),  SGRB hosts and AGN increases with decreasing z (Contini 2017, 2019). Contini (2017) suggested that
 N/O ratios in SN hosts increase due to secondary N production towards low z (0.01). They accompany the growing trend  of
 AGN  and low-ionization nuclear emission-line regions (LINER)(Contini 2018b). 
 N/O ratios in LGRB hosts decrease rapidly between z $>$ 1 and z $\sim$ 0.1 following the N/H trend and reach the 
 characteristic N/O ratios calculated for the HII regions in local and nearby galaxies. 
The  location of the N/O ratio calculated   for the   LGRB031203 host  in  Fig. 6 diagram
reveals that  O/H is relatively low.
The low [NII]/\Hb line ratios in the Margutti et al spectra confirm that N/H is  low in LGRB at low z.
We have found that the low O/H and N/H abundance ratios   in the GRB031203 host  are accompanied by
low  Ne/H, S/H and Ar/H. O, N, Ne, S and Ar  are among the most abundant heavy  elements.
Moreover,  the  fit of the  calculated He lines  could improve adopting  He/H=0.08 (Tables 2 and 3)
in agreement with Guseva et al (2011).
 We will be able to confirm whether more  elements are depleted in LGRB host galaxies   when  C, Si,  Mg, etc  lines
which are strong in the UV and in the IR will be available from the observations. 

Modelling the SED of the GRB031203 host, we have found that the underlying old star population emerges from the 
bremsstrahlung by a black-body bump  in the IR  corresponding to a temperature of 10$^4$K. 
The thermal bremsstrahlung and the synchrotron radiation created by the Fermi mechanism at the shock front
which appear in the SED radio range, belong to different observations.

\begin{figure}
\centering
\includegraphics[width=9.5cm]{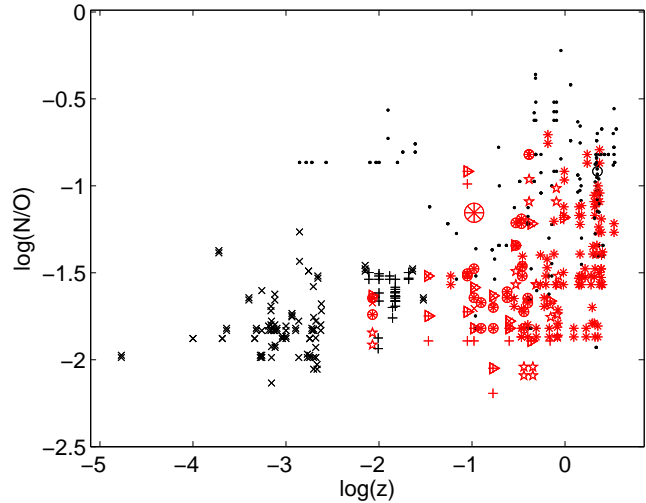}
\caption{N/O abundance ratio results as function of the redshift.
Large encircled red asterisk: N/O calculated for GRB031203. Other symbols
are described in Table 5.
}
\end{figure}

\begin{table}
\centering
\caption{Symbols in Fig. 6}
\begin{tabular}{llcl} \hline  \hline
\ symbol   &object                       & Ref. \\ \hline
\ red  asterisks &LGRB hosts                   & (1) \\
\ red  triangle+cross    & LGRB hosts         & (2)  \\
\ red  pentagrams  & LGRB  hosts     &  (3)  \\
\ red triangle +plus& LGRB hosts with WR stars& (4) \\
\ red encircled asterisks &LGRB at low z     & (5)\\
\ black  dots & star-burst galaxies              & (6)\\
\ black plus &  low-luminosity nearby galaxies & (7)  \\
\ black cross& HII regions in local galaxies   & (8)  \\ \hline
\end{tabular}

(1) Kr\"{u}hler et al. (2015);
(2) Savaglio et al. (2009);
(3) Contini (2016, table 8);
(4) Han et al.(2010);
(5) Niino et al.(2016);
(6) reported by Contini (2014) and references therein;
(7) Marino et al. (2013);
(8) Berg   et al. (2012);
\end{table}

\section*{Acknowledgements}  
I am grateful to the referee for many helpful remarks.
                       
\section*{References}

\def\ref{\par\noindent\hangindent 18pt}

\ref Aldrovandi, S.M.V. \& P\'{e}quignot, D. 1973 A\&A, 25, 137 
\ref Bell, A.R. 1977, MNRAS, 179, 573
\ref Berg, D.A. et al 2012, ApJ, 754, 98
\ref Berger, E. 2009, ApJ, 690, 231
\ref Bloom, J.S., Frail, D.A., Kulkarni, S.R. 2003, ApJ, 591, L13
\ref Chen, H-W. 2018, MNRAS    ArXiv:1110.0487
\ref Contini, M. 2019, submitted
\ref Contini, M. 2018a, arXiv:1801.03312
\ref Contini, M. 2018b, A\&A, 620, 37
\ref Contini, M. 2017, MNRAS, 469, 3125
\ref Contini, M. 2016, MNRAS, 460, 3232
\ref Contini, M. 2014, A\&A, 564, 19
\ref Contini, M. \& Viegas-Aldrovandi, S.M. 1990 ApJ, 350, 125
\ref Contini, M. \& Aldrovandi, S.M.V.  1983 A\&A, 127, 15 
\ref Contini, M. \& Shaviv, G. 1980, A\&A, 88, 117
\ref Cox, D.P. 1972, ApJ, 178, 143
\ref Cox, D.P. \& Tucker, W.H. 1969, ApJ, 157, 1157
\ref Gendre, B., Corsi, A., Piro, L. 2006, A\&A, 455, 803
\ref Crowther, P.A., Dessart, L., Hillier, D.J., Abbott, J.B., Fullertone, A.W. 2002, A\&A, 392, 653
\ref Giannios, D., Mimica, P., Aloy, M.A. 2008, A\&A, 478, 747
\ref Gotz, D., Meneghetti, S.,Beck, M., Borkowski, J. 2003, GCN Circ. 2459
\ref Gotz, D.  Freyer, C.L., Woosley, S.E., Langer, N., Hartmann, D.H. 2003, ApJ, 591, 288 
\ref Guseva, N.G., Izotov, Y.I., Fricke, K.J., Henkel, C. 2011, A\&A, 534, A84
\ref Firmani, C., Ghisellini, G., Avila Reese, V., Ghirlanda, G. 2006, MNRAS, 370, 185
\ref Fruchter, A.S. et al 2006, Nature, 441, 463
\ref Hammer, F., Flores, H., Schaerer, D. et al 2006, A\&A, 454, 103
\ref Han, X. H., Hammer, F., Liang, Y. C., Flores, H., Rodrigues, M., Hou, J. L., Wei, J. Y.  2010, A\&A, 514, 24
\ref Heger, A., Woosley, S.E., Langer, N., Spruit, H.C. 2004, in Stellar rotation, IAU Symp., 215, 591 
\ref Hirschi, R., Meynet, G., Maeder, A. 2005, A\&A, 443, 581
\ref Hjorth, J.J. 2003, Nature, 423, 847
\ref Hsia, C.H., Lin, H.C., Huang, K.J., Urata, Y., Ip, W.H., Tamagawa, T.
2003, GCN Circ. 2470
\ref Kobayashi, S. \& Zhang, B. 2003, ApJ, 597, 455.
\ref Kouveliotou, C., Meegan, C.A., Fishman, G.J et al   1993, ApJ, 413, L101
\ref Kr\"{u}hler, T. et al 2015 A\&A, 581, 125
\ref MacFadyen, A.I. \& Woosley, S.E.  1999, ApJ, 524, 262
\ref Maeder, A, Meynet, G., Hirschi, R. 2005, ASPC, 336, 79
\ref Margutti, R. et al 2008, A\&A, 474, 815
\ref Marino, R.A. et al 2013, A\&A, 229, 114
\ref Mazzali, P.A. et al 2006, ApJ, 645, 1323
\ref Meneghetti, S. \& Gotz, D. 2003, GCN Circ.2460
\ref Niino, Y. et al 2016 Publ. Astron. Soc. Japan, ArXiv:1616.01983
\ref Perley, D.A. , Niino, Y., Tanvir, N.R., Vergani, S.D., Fynbo, J.P.U. 2016,  ApJ ArXiv:1602.00770
\ref Piran, T. 2004 RvMP, 76, 1143 (arXiv:astro-ph/0405503)
\ref Prochaska, J.X. et al 2004, ApJ, 611, 200
\ref Sari, R.\& Piran, T. 1999, ApJ, 520, 641
\ref Savaglio, S., Glazerbrook, K., Le Borgne, D. 2009, ApJ, 691, 182
\ref Schiano, A.V.R. 1985, ApJ, 299, 24
\ref Schiano, A.V.R. 1996, ApJ, 302, 81
\ref Tagliaferri, Covino, S., Fugazza, D., Chincarini, G., Malesani, D., Della Valle, M., Stella, L. 2004IAUC.8308
\ref Thilker, D.A. et al. 2009, Nature, 457, 990
\ref Viegas-Aldrovandi, S.M. \& Contini, M. 1989, A\&A, 215, 253
\ref Viegas, S.M. \& Contini, M. 1994, ApJ, 428, 113
\ref Vink, J.S., de Koter, A. 2005, A\&A, 442, 587
\ref  Vogel, M. \& Nussbaumer, H. 1994, A\&A, 284, 145
\ref Watson, D. et al 2011, ApJ, 741, 58
\ref Wei, J.-J., Wu, X.-F., Melia, F., Wei, D.-M., Feng, L.-L. 2014, MNRAS, 439, 3329
\ref Williams, R.E. 1973, MNRAS, 164, 411
\ref Williams, R.E. 1967, ApJ, 147, 556  
\ref Wolf, C. \& Podsiadlowski, P.    2007, MNRAS, 375, 1049
\ref Woosley, S.E. 1993, ApJ, 405, 237
\ref Woosley, S.E. \& Bloom, J. S. 2006, ARA\&A, 44, 507
\ref Zhang, B. \& Kobayashi, S. 2005, ApJ, 628, 315

\end{document}